\documentclass[preprint,pre,floats,aps,amsmath,amssymb]{revtex4}

\usepackage{graphicx}
\usepackage{bm}
\usepackage[utf8]{inputenc}
\usepackage{upgreek}

\begin{document}

\title{Surfing or sliding: the act of naming and its implications}
\author{M. Hennes, J. Tailleur, Gaelle Charron, and Adrian Daerr$^1$}
\affiliation{Laboratoire Matière et Systèmes Complexes (MSC) UMR CNRS 7057, University Paris Diderot, 75205 Paris cedex 13, France}


\maketitle

Kovàcs et \textit{al.} have greatly contributed to the characterisation of \textit{sliding} \cite{hoelscher2017, grau2015, pollitt2015}, a flagella-independent, "passive type of [bacterial] movement, [...] powered by the pushing force of dividing cells and additional factors facilitating the expansion over surfaces" \cite{hoelscher2017}. They suggest \cite{kovacs2017} that bacterial surfing \cite{hennes2017} should be described by the same name to "facilitate the understanding between the biophysics and the microbiology communities".\\
\\
Providing a clear and systematic classification of bacterial modes of motility is indeed a common objective of both communities, but one still has to agree on the best way to proceed. Bacteria possess a surprisingly rich toolbox to interact with their environment and achieve motion: flagella, pili, surfactants, exopolysaccharide, etc \cite{henrichsen1972}. Because sliding and surfing share overlapping \textit{components}, Kovàcs et \textit{al.} suggest using a unique designation. On the contrary we advocate a characterisation and classification of surface translocation modes by their underlying physical and chemical \textit{mechanisms}. The comparison with human motions speaks for itself: running, swimming, walking and biking all make use of legs, but the diversity of the underlying mechanisms and characteristics has naturally led to an equal diversity of designations. \\
\\ 
Similarly, bacterial surfing and sliding indeed have little in common beyond requiring surfactant production:
\begin{itemize}
\item Sliding is powered by cell division \cite{henrichsen1972, harshey2003, hoelscher2017} while gravity is the driving force of surfing \cite{hennes2017}, the cells being only passengers of the \textit{sliding droplet}.
\item Sliding occurs at high cell density \cite{kinsinger2003} whereas the cell density remains far below close packing in surfing \cite{hennes2017}.
\item Sliding speeds are reported to be typically around 2 $\upmu$m min$^{-1}$ - 5 $\upmu$m min$^{-1}$ \cite{henrichsen1972, harshey2003, seminara2012}, which is about 50 times smaller than typical surfing speeds (250 $\upmu$m min$^{-1}$ even on very shallow slopes).
\item Sliding seems to rely on the production of an EPS matrix \cite{gestel2015}, which is not required in surfing (droplet depinning can be initiated even by pure surfactin or sugar drops \cite{hennes2017}). 
\end{itemize}
For all these reasons, we think using a common name would be a source of confusion, rather than clarity.\\
\\ 
Our characterization of colony surfing has made its underlying mechanisms quite clear: an osmotic pumping makes a droplet containing bacteria inflate while surfactin directly lowers the surface tension and increases the wettability of the substrate, leading to the depinning of the droplet and hence to its sliding. Bacteria are then passively advected by the surrounding fluid, collectively "surfing" on the sliding droplet. \\
\\ 
On the contrary, obtaining a precise characterisation of the microscopic physical and chemical mechanisms underpinning sliding remains a challenge. This is apparent in the recent review of Hölscher and Kovàcs \cite{hoelscher2017} which highlights that "sliding" describes translocation modes that may or may not require surfactant, and may or may not require EPS matrix. Further elucidation of the physico-chemistry of sliding could thus, in our opinion, lead to more precise characterisations of translocation modes which would only phenomenologically appear related: the entropy of our dictionary should only increase as our knowledge progresses. \\
\\ 
Author contributions: MH, JT, GC and AD wrote the manuscript. \\
\\
The authors declare no conflict of interest. \\
\\
$^1$ Corresponding author: \url{adrian.daerr@univ-paris-diderot.fr}


\begin{thebibliography}{99}


\bibitem{hoelscher2017}T. Hölscher, and A.T. Kovàcs (2017) Sliding on the surface: bacterial spreading without an active motor. {\it Environ Microbiol} (in press) doi: 10.1111/1462-2920.13741.

\bibitem{grau2015}R.R Grau, P. de Ona, M. Kunert, C Lenini, R. Gallegos-Monterrosa, E. Mathre, D. Vileta, V. Donato, T. Hölscher, W. Boland, O.P. Kuipers, and A.T. Kovàcs (2015) A duo of potassium-responsive histidine kinases govern the multicellular destiny of Bacillus subtilis. {\it mBio} 6(4):e00581.

\bibitem{pollitt2015}E.J.G. Pollitt, S.A. Crusz, and S.P. Diggle (2015) Staphylococcus aureus forms spreading dendrites that have characteristics of active motility. {\it Scient Rep} 5:17698.

\bibitem{kovacs2017}A.T. Kovàcs, R. Grau, and E. Pollitt (2017) Surfing of bacterial droplets: Bacillus subtilis sliding revisited. {\it Proc Natl Acad Sci USA}.

\bibitem{hennes2017}M. Hennes, J. Tailleur, G. Charron, and A. Daerr (2017) Active depinning of bacterial droplets: the collective surfing of Bacillus subtilis. {\it Proc Natl Acad Sci USA} 114:5958-5963.

\bibitem{henrichsen1972}J. Henrichsen (1972) Bacterial surface translocation: a survey and a classification. {\it Bacteriol Rev} 36(4):478-503.

\bibitem{harshey2003}R.M. Harshey (2003) Bacterial motility on a surface: many ways to a common goal. {\it Annu Rev Microbiol} 57:249-273.

\bibitem{kinsinger2003}R.F. Kinsinger, M.C. Shirk, and R. Fall (2003) Rapid surface motility in Bacillus subtilis is dependent on extracellular surfactin and potassium ion. {\it J Bacteriol} 185(18)5627-5631.

\bibitem{seminara2012}A. Seminara, T.E. Angelini, J.N. Wilking, H. Vlamakis, S. Ebrahim, R. Kolter, D.A. Weitz, and M.P. Brenner (2012) Osmotic spreading of Bacillus subtilis biofilms driven by an extracellular matrix. {\it Proc Natl Acad Sci USA} 109(4):1116-1121.

\bibitem{gestel2015}J. van Gestel, H. Vlamakis, and R. Kolter (2015) From cell differentiation to cell collectives: Bacillus subtilis uses division of labor to migrate. {\it PLoS Biol} 13(4):e1002141. 

\end{thebibliography}
\end{document}